\documentclass[
    ,final            
  ]
  {aipproc}

\layoutstyle{6x9}

\newcommand{\mpi}{m_\pi}

\newcommand{\gev}{\,{\rm GeV}}

\newcommand{\fm}{\,{\rm fm}}
\newcommand{\CC}{{C}}

\begin{document}

\title{Aspects of SU(3) baryon extrapolation}


\author{R.~D.~Young}{
  address={Physics Division, Argonne National Laboratory, Argonne, Illinois 60439, USA}
}

\begin{abstract}
  We report on a recent chiral extrapolation, based on an SU(3)
  framework, of octet baryon masses calculated in 2+1-flavour lattice
  QCD. Here we further clarify the form of the extrapolation, the
  estimation of the infinite-volume limit, the extracted low-energy
  constants and the corrections in the strange-quark mass.
\end{abstract}

\maketitle

%
In recent times, there has been significant advances in numerical
studies of lattice-regularised QCD. In particular, the past year has
seen several studies of baryon systems using 3 flavours of light
dynamical quarks, including
Refs.~\cite{WalkerLoud:2008bp,Aoki:2008sm,Lin:2008pr,Durr:2009ma,Beane:2009ky,Yamazaki:2009zq}.
While the quark masses are relatively light, one still requires an
extrapolation to the physical point to make comparison with
observables. Here we highlight some features of a recent SU(3) chiral
extrapolation of octet baryon masses \cite{Young:2009zb}.

%
The chiral expansion of baryon masses have been studied extensively,
eg. see \cite{Borasoy:1996bx,Donoghue:1998bs}.  For our formulae, we
most closely follow the work of Walker-Loud \cite{WalkerLoud:2004hf},
writing
\begin{equation}
M_B = M^{(0)} + \delta M_B^{(1)} + \delta M_B^{(3/2)} + \ldots\,.
\label{eq:MB32}
\end{equation}
The leading term, $M^{(0)}$, denotes the degenerate mass of the baryon
octet in the SU(3) chiral limit --- where $m_u=m_d=m_s=0$.  The
leading corrections, as one moves away from this limit, are linear in
the quark mass. Assuming SU(2) symmetry in the light quark sector,
with $m_l=(m_u+m_d)/2$, these leadings corrections can be written as
\begin{equation}
\delta M_B^{(1)} = -C_{Bl}^{(1)} m_l - C_{Bs}^{(1)} m_s\,,
\label{eq:dM1}
\end{equation}
with coefficients $C_{Bq}^{(1)}$ listed in Table~\ref{tab:C1} \cite{WalkerLoud:2004hf}.
We make a convenient substitution by
reexpressing the expansion in terms of $\mpi$ and $m_K$, with $m_l\to \mpi^2/2$
and $m_s\to (m_K^2-\mpi^2/2)$. The coefficients, $\alpha_M$, $\beta_M$
and $\sigma_M$ are then each redefined by a dimensionful scale factor,
as similarly done in Ref.~\cite{WalkerLoud:2008bp}. To the order we
work in this manuscript, the use of either quark masses or meson
masses squared is equivalent.

Beyond this linear order come the leading loop corrections, which have
long been known to generate large corrections (at the physical
quark masses), up to ${\cal O}(100\%)$ of the leading terms
\cite{Borasoy:1996bx,Donoghue:1998bs}. This problem in the SU(3)
expansion has been overcome through introducing a finite scale into
the regularization of these loop diagrams
\cite{Donoghue:1998bs}. Concurrently, the introduction of a finite
regularization scale was demonstrated to tame the chiral extrapolation
problem for lattice QCD \cite{Leinweber:1999ig}. While these early
studies required the regularization scale to be input from
phenomenology, this study uses the lattice results themselves to pick
this scale.

We summarize the loop contributions by
\begin{equation}
\delta M_B^{(3/2)} = 
-\frac{1}{16\pi f^2} \sum_\phi\left[ 
\chi_{B\phi} I_R(m_\phi,0,\Lambda) 
+ \chi_{T\phi}I_R(m_\phi,\delta,\Lambda) \right]\,,
\label{eq:dM32}
\end{equation}
where the loop integrals, $I$, and appropriate renormalisations, in a
variety of regularisations, can be found in Ref.~\cite{Young:2002ib}.

The relevant coefficients for these diagrams are displayed in
Table~\ref{tab:chi} \cite{WalkerLoud:2004hf}.  For this study, the
baryon-baryon-meson coupling constants are fixed by phenomenology,
with $D+F=g_A=1.27$ and from SU(6) we use $F=\frac23 D$ and
$\CC=-2D$. We note a similar value for $\CC$ can also be inferred from
the decay width of the $\Delta$.  We adopt a chiral perturbation
theory estimate for the pion decay constant in the SU(3) chiral limit,
$f=0.0871\gev$ \cite{Amoros:2001cp}. The final phenomenological input
we use is the octet-decuplet splitting, where we take the physical
$N$--$\Delta$ splitting, $\delta=0.292\gev$. All these inputs could
potentially be constrained by actual lattice simulation results, or at
least in the near future.  The full analysis reported in
\cite{Young:2009zb} allows each of the chiral axial charges to vary
from these estimates by 15\%, and a 5\% variation in $f$ ---
precisions which should certainly be within reach of the current or
next generation of simulations.

\begin{table}
  \caption{Coefficients for the leading quark-mass expansion of the octet baryons 
    about the SU(3) chiral limit.
  \label{tab:C1}}
\begin{tabular}{lcc}
\hline
$B$       & $C_{Bl}^{(1)}$ & $C_{Bs}^{(1)}$ \\
\hline
$N$       & $2\alpha_M+2\beta_M+4\sigma_M$             & $2\sigma_M$ \\
$\Lambda$ & $\alpha_M+2\beta_M+4\sigma_M$              & $\alpha_M+2\sigma_M$ \\
$\Sigma$  & $\frac53\alpha_M+\frac23\beta_M+4\sigma_M$ & $\frac13\alpha_M+\frac43\beta_M+2\sigma_M$ \\
$\Xi$     & $\frac13\alpha_M+\frac43\beta_M+4\sigma_M$ & $\frac53\alpha_M+\frac23\beta_M+2\sigma_M$ \\
\hline
\end{tabular}
\end{table}

\begin{table}
\begin{tabular}{l|ccc|ccc}
\hline
           & \multicolumn{3}{c|}{$\chi_{B\phi}$} & \multicolumn{3}{c}{$\chi_{T\phi}$} \\
           & $\pi$ & $K$ & $\eta$ & $\pi$ & $K$ & $\eta$ \\
\hline
$N$       & $\frac32(D+F)^2$ & $\frac13(5D^2-6DF+9F^2)$ & $\frac16(D-3F)^2$ & $\frac43\CC^2$ & $\frac13\CC^2$ & $0$ \\
$\Lambda$ & $2D^2$ & $\frac23(D^2+9F^2)$ & $\frac23 D^2$ & $\CC^2$ & $\frac23\CC^2$ & $0$ \\
$\Sigma$  & $\frac23(D^2+6F^2)$ & $2(D^2+F^2)$ & $\frac23 D^2$ & $\frac29\CC^2$ & $\frac{10}{9}\CC^2$ & $\frac13\CC^2$ \\
$\Xi$     & $\frac32(D-F)^2$ & $\frac13(5D^2+6DF+9F^2)$ & $\frac16(D+3F)^2$ & $\frac13\CC^2$ & $\CC^2$ & $\frac13\CC^2$ \\
\hline
\end{tabular}
\caption{The relevent coefficients for the one-loop diagrams 
involving intermediate octet and decuplet baryons. These contributions 
are counted at the same chiral order in the present scheme.
  \label{tab:chi}}
\end{table}

%
We apply our chiral expansion to two recent 2+1-flavour calculations
of the octet baryon spectrum, those by LHPC \cite{WalkerLoud:2008bp}
and PACS-CS \cite{Aoki:2008sm}. To determine the absolute masses of
the baryons, we need to rely on the lattice scale to be fixed from an
alternate source. The MILC collaboration have gone to tremendous
effort to tune the lattice spacing to reproduce hyperfine splittings
in $b$-quark systems \cite{Aubin:2004wf}.  For the purposes of
matching, it is convenient to relate to this scale determination
through the Sommer scale, using $r_0=0.465\pm 0.012\fm$
\cite{Aubin:2004wf}.

\begin{figure}
\includegraphics[width=0.44\columnwidth,angle=0]{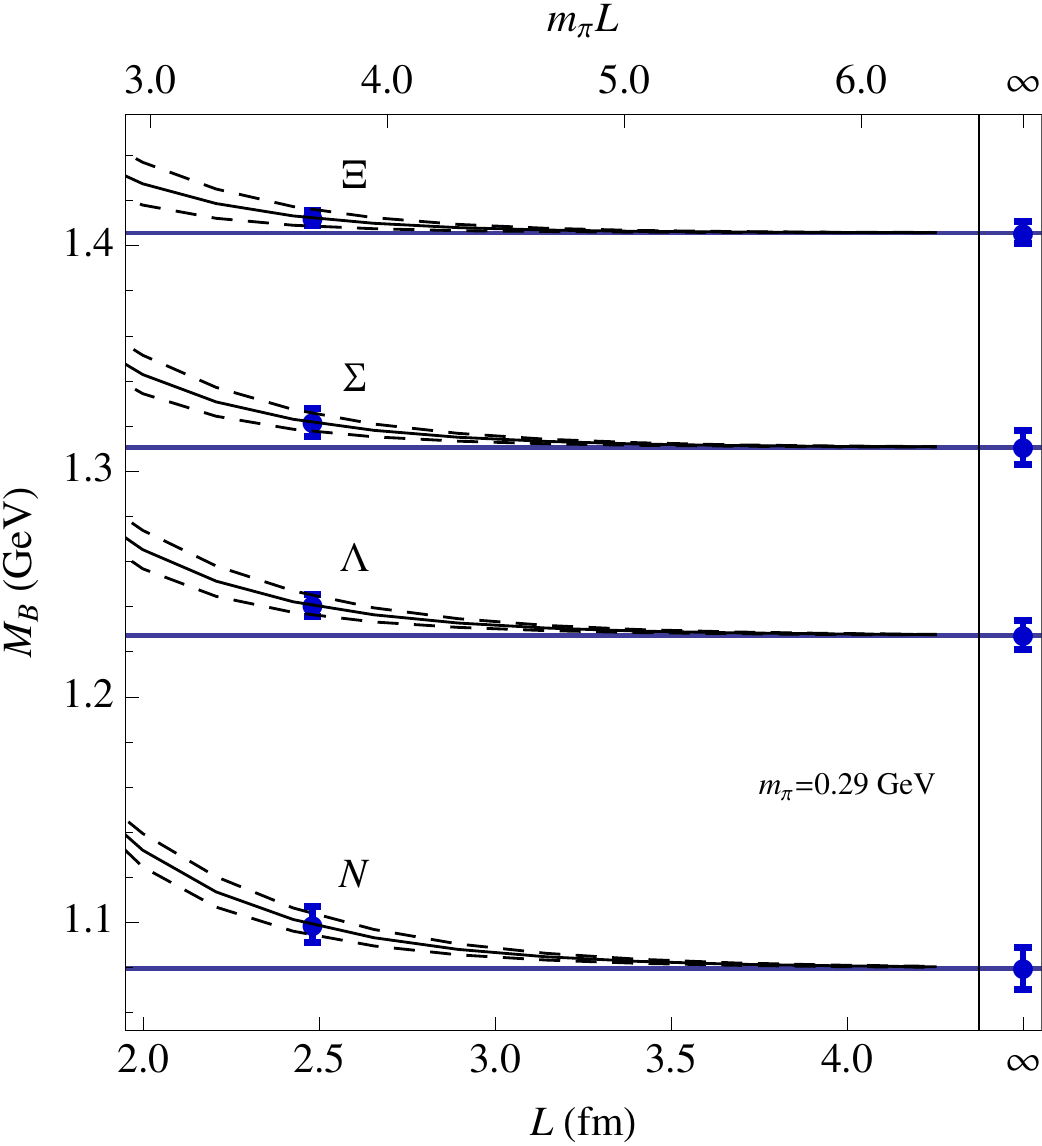}
\hspace*{3mm}
\includegraphics[width=0.44\columnwidth,angle=0]{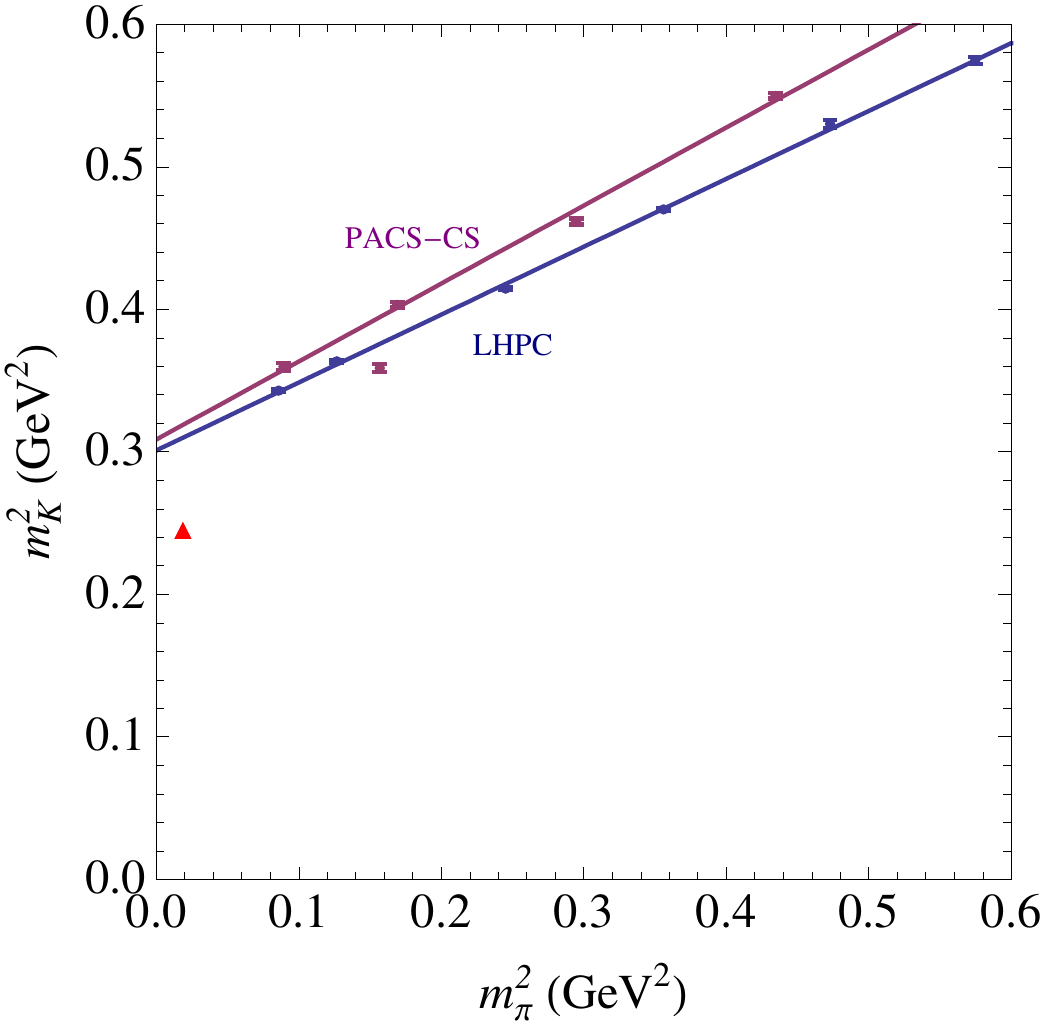}
\caption{(left panel) The baryon masses are shown for the lightest
  pion mass of LHPC simulations. The dashed bands denote an estimate of
  the uncertainty in the finite-volume correction, as described in the
  text.  The horizontal lines depict the infinite volume limit, with
  the combined uncertainty shown by the data point at the right edge
  of the graph.  (right panel) Simple linear extrapolations of the
  kaon mass squared for the PACS-CS and LHPC results. The physical kaon mass is denoted by the triangle.
  \label{fig1}}
\end{figure}

The expansion described is formulated for the continuum and
infinite-volume limit. While the approach to the continuum can be
improved by choice of lattice action, the finite-volume effects are
dominated by infrared chiral physics. This is the same chiral physics
that contributes to the chiral expansion described above, where the
leading finite-volume effects can be estimated by replacing the
continuum loop integrals with discrete sums over lattice momenta
\cite{Young:2002cj,AliKhan:2003cu,Beane:2004tw}.  In the left panel of
Figure~\ref{fig1}, we show an example of the EFT prediction of the
approach to the infinite-volume limit --- a prediction that is
strongly supported by numerical studies in SU(2) simulations
\cite{AliKhan:2003cu}. Our band shows potential differences in how the
ultraviolet component of the meson loops are treated, using either a
finite regulator \cite{Young:2002cj} or a scale-free approach
\cite{Beane:2004tw}. The uncertainty is added in quadrature to the
statistical error in the estimate of the infinite-volume limit.

\begin{figure}
\includegraphics[width=0.62\columnwidth,angle=0]{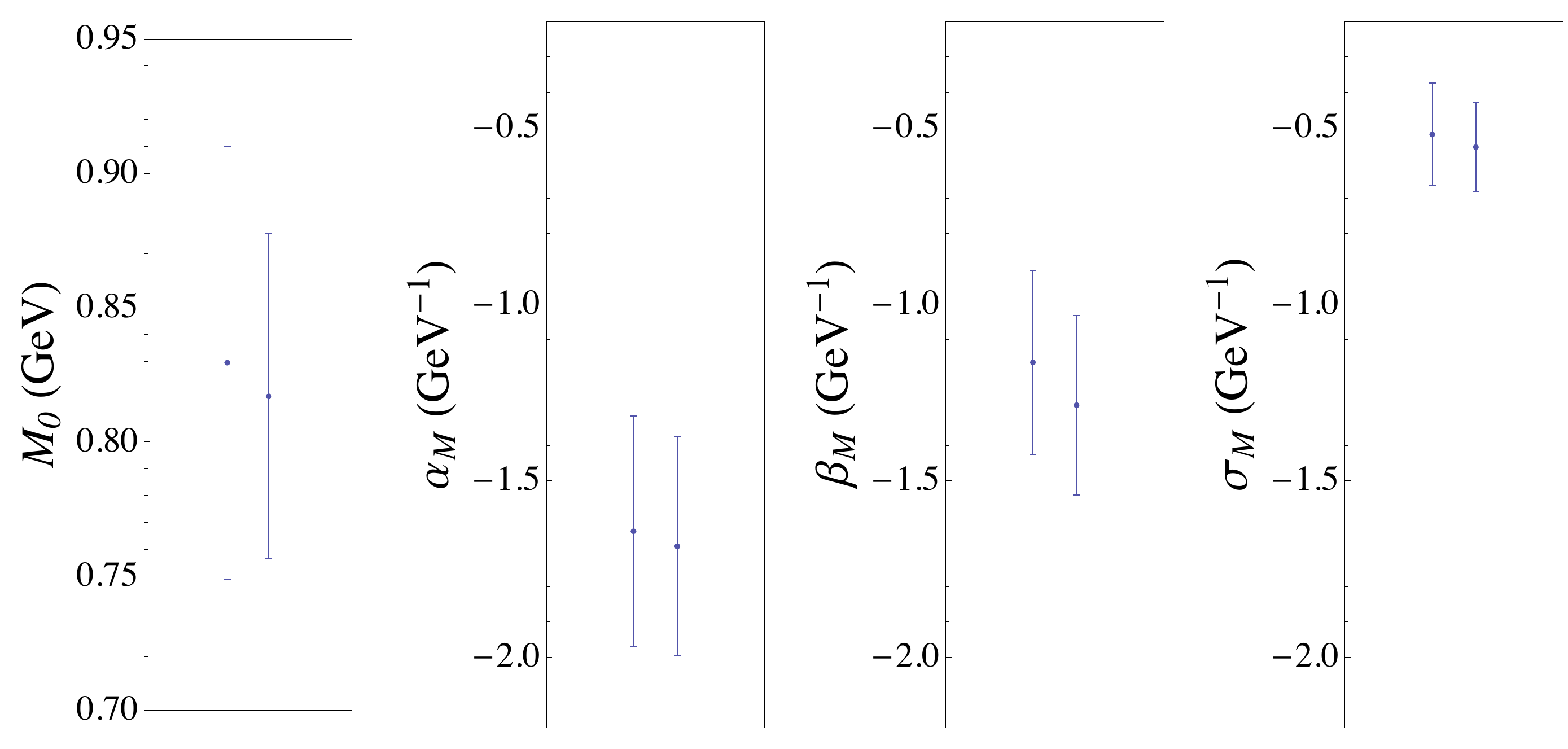}
\caption{Best-fit parameters for the fits to the PACS-CS (left)
and LHPC (right) lattice results.
\label{fig:lec}}
\end{figure}

We choose to remain as close to the chiral regime as possible, and fit
just those results for $\mpi^2<0.2\gev^2$. Independent fits to each of
the LHPC and PACS-CS data sets yield excellent agreement in the
low-energy constants of the fits, see Fig.~\ref{fig:lec}.
Further, the one additional fit parameter, given by the finite
regularization scale is also in agreement between the two
simulations.

This determination of the LECs and the relevant correlations lead to
the accurate values for the baryon masses reported in
\cite{Young:2009zb}.  Beyond the masses that have been extracted, the
analysis demonstrates the ability to correct for the difference
between the lattice and physical strange quark masses --- in this
case, the larger lattice strange quark masses are realised in the
right panel of Fig.~\ref{fig1}. The reliability of these corrections
is demonstrated in \cite{Young:2009zb}, where the fit to the PACS-CS
results are used to predict the baryon masses of the different strange
quark mass ensemble.

Expressions for the baryon masses as a function of the meson masses
(or quark masses by Gell-Mann--Oakes--Renner) have also enabled
accurate determinations of the relevant sigma terms
\cite{Young:2009zb}.  These sigma terms are of particular importance
in the context of current dark matter searches \cite{Ellis:2008hf}.
Our extracted values are in satisfactory agreement with both the
latest light-quark sigma term inferred from experiment
\cite{Pavan:2001wz}, and other recent lattice determinations of the
strange-quark sigma term \cite{Ohki:2008ff,Toussaint:2009pz}.  These
``small'' strangeness sigma terms are noted to be consistent with
phenomenological expectations \cite{Flambaum:2004tm,Chang:2009ae}.


I thank A.~W.~Thomas for collaboration on the work presented here.
This work was supported by DOE contracts 
DE-AC02-06CH11357, under which UChicago Argonne, LLC, operates Argonne
National Laboratory.

\end{document}